\documentstyle[12pt]{article}
\makeatletter

\newcommand{\Classification}[1]{\let\@thefnmark\relax
\@footnotetext{{\bf 1991 Mathematical Subject Classification: } #1}}

\@addtoreset{equation}{section}

\def\section#1{\@ifstar{Nsection*}{\def\thesection{\arabic{section}.}\@startsection{section}{1}{\z@}{-3.5ex
plus -1ex minus
 -.2ex}{2.3ex plus .2ex}{\bf}{\uppercase\expandafter{#1}}
\def\thesection{\arabic{section}}}}

\def\subsection{\@startsection{subsection}{2}{\z@}{-3.25ex plus -1ex minus
 -.2ex}{-1em}{\bf}}

\def\thebibliography#1{\par\vskip 6 mm\noindent {\bf REFERENCES} \par \list
 {[\arabic{enumi}]}{\settowidth\labelwidth{[#1]}\leftmargin\labelwidth
 \advance\leftmargin\labelsep
 \usecounter{enumi}}
 \def\newblock{\hskip .11em plus .33em minus -.07em}
 \sloppy
 \sfcode`\.=1000\relax}

\tabbingsep \labelsep

\def\address#1{\gdef\@address{#1}}

\def\maketitle{\par
 \begingroup
 \def\thefootnote{\fnsymbol{footnote}}
 \def\@makefnmark{\hbox
 to 0pt{$^{\@thefnmark}$\hss}}
 \if@twocolumn
 \twocolumn[\@maketitle]
 \else \newpage
 \global\@topnum\z@ \@maketitle \fi\thispagestyle{empty}\@thanks
 \endgroup
 \setcounter{footnote}{0}
 \let\maketitle\relax
 \let\@maketitle\relax
 \gdef\@thanks{}\gdef\@author{}\gdef\@title{}\let\thanks\relax}
\def\@maketitle{\newpage
 \null
 \vskip 10mm \begin{center}
 {\bf\uppercase\expandafter{\@title}\par} \vskip 4 mm 
 {\bf\@author \par}
 \end{center}
 \vskip 20 mm}

\newcommand{\nc}{\newcommand}
\newtheorem{lemma}{Lemma} [section]
\newtheorem{theorem}[lemma]{Theorem}
\newtheorem{proposition}[lemma]{Proposition}
\newtheorem{corollary}[lemma]{Corollary}
\newtheorem{remark}[lemma]{Remark}
\newtheorem{assumption}[lemma]{Assumption}
\newtheorem{example}[lemma]{Example}

\newtheorem{definition}[lemma]{Definition}

\nc{\QED}{\mbox{}\hfill \raisebox{-2pt}{\rule{5.6pt}{8pt}\rule{4pt}{0pt}} 
          \medskip\par}

\nc{\di}{\displaystyle}
\nc{\SS}{{\rm I\mkern-4.0mu S}}
\nc{\SSs}{{\sf S\mkern-6.5mu S}} 
\nc{\C}{{\rm I\mkern-4.0mu C}}        
\nc{\R}{{\rm I\mkern-4.0mu R}}
\nc{\Z}{{\sf Z\mkern-6.5mu Z}}
\nc{\CCa}{\mbox{$C \mkern-10.5mu\raisebox{.18em}{$\scriptstyle/$}\mkern3mu$}}
\nc{\CCb}{C\mkern-11.5mu I\mkern4mu}
\nc{\Cc}{\mathchoice{\CCa}{\CCa}{\CCb}{C\mkern-10.5mu I\mkern3mu}}
\renewcommand{\Re}{\mathop{\rm Re}}

\nc{\Ref}[1]{~\mbox{\rm (\ref{#1})}}       
\nc{\REF}[1]{~\mbox{\rm \ref{#1}}}             
\nc{\Norm}[2]{\|#1\|\left.\vphantom{T_{j_0}^0}\!\!\right._{#2}}                    
                                               
\nc{\Log}{\mathop {\rm Log}\nolimits}

\expandafter\chardef\csname pre amssym.def at\endcsname=\the\catcode`\@
\catcode`\@=11

\def\undefine#1{\let#1\undefined}
\def\newsymbol#1#2#3#4#5{\let\next@\relax
 \ifnum#2=\@ne\let\next@\msafam@\else
 \ifnum#2=\tw@\let\next@\msbfam@\fi\fi
 \mathchardef#1="#3\next@#4#5}
\def\mathhexbox@#1#2#3{\relax
 \ifmmode\mathpalette{}{\m@th\mathchar"#1#2#3}%
 \else\leavevmode\hbox{$\m@th\mathchar"#1#2#3$}\fi}
\def\hexnumber@#1{\ifcase#1 0\or 1\or 2\or 3\or 4\or 5\or 6\or 7\or 8\or
 9\or A\or B\or C\or D\or E\or F\fi}

\font\tenmsa=msam10
\font\sevenmsa=msam7
\font\fivemsa=msam5
\newfam\msafam
\textfont\msafam=\tenmsa
\scriptfont\msafam=\sevenmsa
\scriptscriptfont\msafam=\fivemsa
\edef\msafam@{\hexnumber@\msafam}
\mathchardef\dabar@"0\msafam@39
\def\dashrightarrow{\mathrel{\dabar@\dabar@\mathchar"0\msafam@4B}}
\def\dashleftarrow{\mathrel{\mathchar"0\msafam@4C\dabar@\dabar@}}

\def\ulcorner{\delimiter"4\msafam@70\msafam@70 }
\def\urcorner{\delimiter"5\msafam@71\msafam@71 }
\def\llcorner{\delimiter"4\msafam@78\msafam@78 }
\def\lrcorner{\delimiter"5\msafam@79\msafam@79 }
\def\yen{{\mathhexbox@\msafam@55 }}
\def\checkmark{{\mathhexbox@\msafam@58 }}
\def\circledR{{\mathhexbox@\msafam@72 }}
\def\maltese{{\mathhexbox@\msafam@7A }}

\font\tenmsb=msbm10
\font\sevenmsb=msbm7
\font\fivemsb=msbm5
\newfam\msbfam
\textfont\msbfam=\tenmsb
\scriptfont\msbfam=\sevenmsb
\scriptscriptfont\msbfam=\fivemsb
\edef\msbfam@{\hexnumber@\msbfam}
\def\Bbb#1{\fam\msbfam\relax#1}
\def\widehat#1{\setbox\z@\hbox{$\m@th#1$}%
 \ifdim\wd\z@>\tw@ em\mathaccent"0\msbfam@5B{#1}%
 \else\mathaccent"0362{#1}\fi}
\def\widetilde#1{\setbox\z@\hbox{$\m@th#1$}%
 \ifdim\wd\z@>\tw@ em\mathaccent"0\msbfam@5D{#1}%
 \else\mathaccent"0365{#1}\fi}
\font\teneufm=eufm10
\font\seveneufm=eufm7
\font\fiveeufm=eufm5
\newfam\eufmfam
\textfont\eufmfam=\teneufm
\scriptfont\eufmfam=\seveneufm
\scriptscriptfont\eufmfam=\fiveeufm

\let\goth\frak

\catcode`\@=\csname pre amssym.def at\endcsname

\begin{document}

\title{The discrete spectrum in the singular Friedrichs model}
\author{ D.Yafaev
\\Universit\'e de Rennes}
\maketitle

\Classification{ Primary 35J10, 47A75; Secondary 81U20}

\vspace{-15mm}
 
\centerline{\bf Dedicated to  M. S. Birman on the occasion of his seventieth birthday}

\vspace{5mm}

\begin{abstract}
A typical result of the paper is the following.  Let ${\bf H}_\gamma={\bf H}_0+\gamma {\bf V}$ where
${\bf H}_0$ is multiplication by
$|x|^{2l}$ and ${\bf V}$ is an integral operator with kernel $\cos\langle  x,y\rangle$ in the space
$L_2({\Bbb R}^d)$. If $l=d/2+ 2k$ for some $k= 0,1,\ldots$, then the operator ${\bf H}_\gamma$ has
infinite number of negative eigenvalues for any coupling constant $\gamma\neq 0$.
 For other values of $l$, the negative spectrum of  ${\bf H}_\gamma$ is infinite for
  $|\gamma|> \sigma_l$ where $\sigma_l$ is some explicit positive
constant. In the  case $\pm \gamma\in (0,\sigma_l]$, the number ${\bf N}^{(\pm)}_l$ of negative
eigenvalues of 
${\bf H}_\gamma$ is   finite  and does not depend on $\gamma$. We  calculate ${\bf N}^{(\pm)}_l$.

\end{abstract}

\section {Introduction}

A perturbation of a multiplication operator $H_0$ by an integral operator $V$ is called a
Friedrichs model. It is usually  assumed that kernel of $V$ is a H\"older continuous
(with an exponent larger than $1/2$) function ${\bf v}(x,y)$ of one-dimensional variables $x$ and
$y$ which decays sufficiently rapidly as
$|x|+|y|\rightarrow\infty$. Then (see \cite{F}) the wave operators for the pair $H_0$, $H_1=H_0+V$
 exist  and are complete. Moreover, the operator $H_1$ does not have
 the singular continuous spectrum, and its discrete spectrum is finite.
It is important that the results of \cite{F} are applicable to the case when  a kernel ${\bf
v}(x,y)$ is itself a compact operator in an auxiliary Hilbert space.

A somewhat different situation was considered in the paper \cite{BY} where the operator ${\bf H}_0$
of multiplication  by $|x|^{2l},\; l>0,$ in the space $L_2({\Bbb R}^d)$ was  perturbed by an integral
operator of Fourier type. More precisely, the perturbation was defined by one of the equalities
\begin{equation}
({\bf V}^{(c)}f)(x)=(2\pi)^{-d/2}\int_{{\Bbb R}^d}\cos \langle x,y\rangle f(y) dy \quad{\rm or}\quad
({\bf V}^{(s)}f)(x)=(2\pi)^{-d/2}\int_{{\Bbb R}^d}\sin \langle x,y\rangle f(y) dy
\label{eq:I1}\end{equation}
 and ${\bf H}_\gamma^{(c)}={\bf H}_0+\gamma {\bf V}^{(c)}$ or ${\bf H}_\gamma^{(s)}={\bf H}_0+\gamma
{\bf V}^{(s)}$ where
$\gamma\in{\Bbb R}$ is a coupling constant. An interesting feature of this model is that
${\bf V}^{(c)}$ and ${\bf V}^{(s)}$ are invariant with respect to the Fourier transform so that one
could have chosen
${\bf H}_0=(-\Delta)^l$ for the ``unperturbed" operator. Passing to the spherical coordinates and
considering the space
$L_2({\Bbb R}^d)$ as $L_2({\Bbb R}_+; L_2({\Bbb S}^{d-1}))$, we can fit the operators
${\bf H}_\gamma^{(c)}$ and ${\bf H}_\gamma^{(s)}$ into the framework of the Friedrichs model.
However, since the kernels $\cos\langle x,y \rangle $ or $\sin\langle x,y \rangle $ do not tend to
$0$ as
$|x|\rightarrow\infty$ or (and) $|y|\rightarrow\infty$, the results of the paper \cite{F} are not
applicable to perturbations (\ref{eq:I1}) (even in the case $d=1$).

Due to oscillations of its kernel, the operators (\ref{eq:I1}) are bounded 
in the space $L_2({\Bbb R}^d)$ but they are not compact, even  relatively with respect to
${\bf H}_0$. Nevertheless, as shown in \cite{BY}, the essential spectra of the operators  ${\bf
H}_\gamma^{(c)}$, ${\bf H}_\gamma^{(s)}$  and ${\bf H}_0$ coincide. This implies that the negative
spectra of the operators
${\bf H}_\gamma^{(c)}$ and ${\bf H}_\gamma^{(s)}$ consist of eigenvalues of finite multiplicity
which may accumulate at the bottom of the essential spectrum  (point zero) only. Moreover, in the
case
$2l>d$, the trace-class technique was used in
\cite{BY} to prove, for the pairs ${\bf H}_0$, ${\bf H}_\gamma^{(c)}$ and ${\bf H}_0$, ${\bf
H}_\gamma^{(s)}$, the existence of the wave operators and their completeness.

Our goal here is to study the discrete spectrum of the operators ${\bf H}_\gamma^{(c)}$ and ${\bf
H}_\gamma^{(s)}$. The space $L_2({\Bbb R}^d)$ decomposes into the orthogonal sum of subspaces
${\goth H}_n,\; n=0,1,2,\ldots$ constructed in terms of the spherical functions of order $n$ and
invariant with respect to the operators ${\bf H}_\gamma^{(c)}$ and ${\bf H}_\gamma^{(s)}$. On the
subspaces ${\goth H}_n$, these operators reduce to operators $H_\gamma $ acting in the
space $L_2({\Bbb R}_+)$. Operators $H_\gamma $ have a form
$H_\gamma =H_0+\gamma V$
where $H_0$ is again multiplication  by
$x^{2l},\; l>0,$ and
\begin{equation} 
 (Vf)(x)=\int_0^\infty v (xy) f(y) dy.
\label{eq:v1}\end{equation}
The function $v$ depends of course on $n$ and on the index $``c"$ or $``s"$ but is always  
expressed in terms of a Bessel function.

Therefore, we consider first operators
$H_\gamma =H_0+\gamma V$ in the space $L_2({\Bbb R}_+)$ where $V$ is given by (\ref{eq:v1})
 with  a rather arbitrary real function $v$.
We suppose that $v(t)$ has a sufficiently regular behaviour as $t\rightarrow 0$ and
$t\rightarrow\infty$. Since operators (\ref{eq:v1}) never satisfy conditions of \cite{F}, this
version of the Friedrichs model is called singular here. 
  It turns out that the negative spectrum of the operator $H_\gamma $ is very sensitive with respect
to the coupling constant $\gamma$ and the parameter $l$. Thus, if the asymptotic expansion of $v(t)$
as $t\rightarrow 0$ contains the term $v  t^r$, then, in the case $l=r +1/2$,  
the operator $H_\gamma$ has infinite number of negative eigenvalues for any coupling constant
$\gamma\neq 0$. For other values of $l$, the negative spectrum of  $H_\gamma$ is infinite for
 $|\gamma|> \sigma_l$ where $\sigma_l$ is some explicit positive constant.
In the  case $\pm\gamma\in(0,\sigma_l]$,  the number $N_l^{(\pm)}$ of negative eigenvalues of 
$H_\gamma$ does not depend on $\gamma$. We  calculate  $N_l^{(\pm)}$ in terms of the asymptotic
expansion of the function $v(t)$ as $t\rightarrow 0$. We emphasize that both positive and negative
parts of perturbation (\ref{eq:v1}) are, in general, non-trivial, and our results on the discrete
spectrum of $H_\gamma$ take into account their ``interaction".

The results  on the discrete spectrum in the singular Friedrichs model can be compared with those
for the  Schr\"odinger operator $-\Delta+\gamma q(x)$ whose potential $q(x)$ has a critical decay at
infinity. Suppose, for example, that $x\in{\Bbb R}$ and a negative function $q(x)$ has the
asymptotics
$q(x)\sim -|x|^{-2}$ as $|x|\rightarrow\infty$. Then for sufficiently small $\gamma >0$ the
negative spectrum of the operator $-\Delta+\gamma q(x)$ consists of exactly one eigenvalue, it is
finite for $\gamma\in (0,1/4)$ and it is infinite for $\gamma > 1/4$.
 Note also that the mechanism for appearance of 
 infinite number of  negative eigenvalues in the singular Friedrichs model  resembles the same
phenomena (the Efimov's effect, see\cite{Y}) for the three-particle Schr\"odinger operator with
short-range pair potentials.

Our calculation of the number of negative eigenvalues relies on the Birman-Schwinger principle. We
need a presentation of this theory somewhat different from the original paper \cite{B} and
adapted to perturbations without a definite sign. This is done in Section 2 in the abstract
framework. The negative spectrum of the operator $H_\gamma$ in the space $L_2({\Bbb R}_+)$ 
is studied in Section 3. In Section 4, we use these results to calculate the total number of negative
eigenvalues of the operators ${\bf H}_\gamma^{(c)}$ and ${\bf H}_\gamma^{(s)}$ in the space
$L_2({\Bbb R}^d)$.

\section {the Birman-Schwinger principle}

{\bf 1.}
Let $H_0$ be an arbitrary self-adjoint   positive operator with domain
 ${\cal D}(H_0)$ in a Hilbert space  ${\cal H}$. Following \cite{B}, we define the ``full"
Hamiltonian $H=H_0+ V$   by means of the corresponding  quadratic form. We suppose that the real
quadratic form
$v[f,f]$ of the perturbation $V$ is defined for $f\in{\cal D}(H_0^{1/2})$ and is compact in the
Hilbert space ${\cal D}(H_0^{1/2})$ with the scalar product
$(f_1,f_2)_{H_0^{1/2}}=(H_0^{1/2}f_1,H_0^{1/2}f_2) + (f_1,f_2)$. This means that
$|v[f,f]|\leq C ||f||_{H_0^{1/2}}^2$ and the operator $T$ defined by the relation
\begin{equation}
(Tu,u)=v[(H_0+I)^{-1/2}u, (H_0+I)^{-1/2}u],\quad
\forall u\in{\cal H},
\label{eq:BS0}\end{equation}
 is compact in the space ${\cal H}$.
  Under  this assumption  the form
\[
h[f,f]= || H_0^{1/2}f||^2 + v[f,f]
\]
is closed on ${\cal D}(H_0^{1/2})$. So  there exists (see \cite{BS}) a unique  self-adjoint,
semi-bounded from below,  operator $H$ such that 
${\cal D}(|H|^{1/2})={\cal D}(H_0^{1/2})$ and $h[f,g]=(Hf,g)$ for any $f\in D(H)$
 and any $g\in D(H_0^{1/2})$. 

The resolvent identity for the operators $H_0,H$ can be written as
\begin{equation} 
(H+c)^{-1} - (H_0+c)^{-1}=-(H_0+c)^{-1} (H_0+I)^{1/2}T(H_0+I)^{1/2} (H+c)^{-1} ,
\label{eq:RI}\end{equation}
where $c$ is a  sufficienly large positive constant. Since the operator $(H_0+I)^{1/2} (H+c)^{-1}$
is bounded, the right-hand side of (\ref{eq:RI}) is a compact operator, and hence by  Weyl's
theorem,  the essential spectra
$\sigma_{ess}$ of the operators $H_0$ and $H$ coincide. This result was established in \cite{B}. 

Our goal here is to calculate the total multiplicity 
\begin{equation}
N=\dim E_H(-\infty,0)
\label{eq:N}\end{equation}
 of the negative spectrum of the operator $H$. To that end we
introduce some auxiliary objects. Let us consider the set ${\cal R}$ of elements $u=H_0^{1/2}f$ for
all
$f\in {\cal D}(H_0^{1/2})$. The set ${\cal R}$ is endowed with the norm
\begin{equation}
 ||u||_{\cal R}^2=||u||^2 + ||H_0^{-1/2}u||^2
\label{eq:R}\end{equation}
 and is dense in ${\cal H}$ since ${\rm Ker}\, H_0^{1/2}=\{0\}$.   Let us
define the bounded quadratic form $a$ on ${\cal R}$ by the relation
\begin{equation}
 a[u,u]=v[ H_0^{-1/2}u, H_0^{-1/2}u],\quad u \in {\cal R}.
\label{eq:BS1}\end{equation}

\begin{lemma}\label{ab1}
The number $(\ref{eq:N})$ equals
the maximal dimension of subspaces ${\cal L}\subset {\cal R}$ such that 
\begin{equation}
 a[u,u] < -||u||^2,\quad \forall u\in {\cal L}.
\label{eq:abs3}\end{equation}
\end{lemma}
{\it Proof.} --
 By the spectral theorem, $N$ equals the maximal dimension of subspaces ${\cal K}\subset {\cal
D}(H_0^{1/2})$ such that
\begin{equation}
 h[f,f] < 0,\quad \forall f\in {\cal K}.
\label{eq:abs3x}\end{equation}
Setting $u=H_0^{1/2}f$ and taking into account definition (\ref{eq:BS1}), we see that
(\ref{eq:abs3}) and (\ref{eq:abs3x}) are equivalent if ${\cal L}=H_0^{1/2}{\cal K}$. Moreover,
$\dim{\cal K}=\dim{\cal L}$   since ${\rm Ker}\, H_0^{1/2}=\{0\}$.
$\quad\Box$

\begin{lemma}\label{ab2}
 Let $B$ be any self-adjoint operator and let ${\goth M}\subset {\cal D}(B)$ be some set dense in
${\cal D}(B)$ in the $B$-metrics defined by
$ ||u||^2_B= ||B u||^2+||u||^2$.
  Then the total multiplicity $\dim E_B(-\infty,0)$ of the
negative spectrum of the operator $B$ equals the maximal dimension of subspaces ${\cal L}\subset
{\goth M}$ such that 
$ (Bu,u) < 0,\; \forall u\in {\cal L}$.
\end{lemma}

In the case of a bounded operator $B$, this assertion is Lemma~1.2 from \cite{B}. In the general
case, its proof is practically the same.

{\bf 2.}
In terms of the form (\ref{eq:BS1}), a simple version of the Birman-Schwinger principle can be
formulated  as follows.

\begin{proposition}\label{BSs}
Suppose that
\begin{equation}
 a[u,u]=(Au,u)  
\label{eq:aA}\end{equation}
for some bounded self-adjoint operator $A$ in the space ${\cal H}$  and   all $ u\in {\cal R}$. Let
\begin{equation} 
M=\dim E_A(-\infty,-1)
\label{eq:M}\end{equation}
be the total multiplicity  of the  spectrum of the operator $A$ in the interval
$(-\infty,-1)$. Then the numbers $(\ref{eq:N})$ and $(\ref{eq:M})$ are equal, that is $N=M$.
\end{proposition}
{\it Proof.} --
Let $n$ be the maximal dimension of subspaces
${\cal L}_0\subset {\cal R}$ such that 
\begin{equation}
 (Au_0,u_0) < -||u_0||^2,\quad \forall u_0\in {\cal L}_0.
\label{eq:pr4a}\end{equation}
It follows from (\ref{eq:aA}) and Lemma~\ref{ab1} that $N=n$. Since ${\cal R}$ is dense in 
${\cal H}$, the equality $M=n$ follows from Lemma~\ref{ab2}.$\quad\Box$ 

This result is contained in the paper \cite{B} because, under assumptions of 
Proposition~\ref{BSs}, the form $v[f,f]$ is bounded in the space ${\goth
H}$ with the norm 
$|| f||_{\goth H}=|| H_0^{1/2}f||$ and consequently is closable in this space.
A more difficult case when the form $v[f,f]$ is not closable in  ${\goth H}$ was also investigated
in \cite{B}. Our aim here is to reconsider this situation and to formulate results in terms of the 
form (\ref{eq:BS1}). This is convenient in the case when the form $v[f,f]$ takes values of both
signs. Actually, we suppose that equality (\ref{eq:aA}) holds only up to some finite number of
squares of (unbounded) functionals $\varphi_1,\ldots, \varphi_n$, that is
\begin{equation}
 a[u,u]=(Au,u)-\sum_{j=1}^m |\varphi_j(u)|^2 + \sum_{j=m+1}^n |\varphi_j(u)|^2. 
\label{eq:abs1}\end{equation}
Of course, one or both sums in (\ref{eq:abs1}) may be absent, that is we do  not exclude the cases
$m=0$ or (and) $n=m$.

Let us give first sufficient conditions for the negative spectrum of the operator $H$ to be
infinite. In this case we do not assume that $A$ is bounded and replace equality (\ref{eq:abs1}) by
an estimate.

\begin{theorem}\label{abs1}
 Let $A$ be a self-adjoint operator with domain ${\cal D}(A)$. Suppose  that a linear set  ${\cal
R}_0\subset {\cal D}(A)\cap {\cal R}$   is dense in ${\cal D}(A)$ in the $A$-metrics.
 Assume that for all $u\in {\cal R}_0$ the form $(\ref{eq:BS1})$ satisfies the estimate
\begin{equation}
 a[u,u] \leq (Au,u) + \sum_{j=1}^p |\varphi_j(u)|^2,
\label{eq:abs0}\end{equation}
where  $\varphi_1,\ldots,\varphi_p$ is a system of linear   functionals defined on ${\cal R}_0$. 
Then the negative spectrum of the operator $H$ is infinite provided 
$\dim E_A(-\infty,-1)=\infty$.
\end{theorem}
{\it Proof.} --
By Lemma~\ref{ab2}, for any $k$, there exists a subspace ${\cal L}_0 \subset {\cal R}_0$ such that
$\dim{\cal L}_0=k$ and  inequality (\ref{eq:pr4a}) holds. Let  ${\cal L}\subset {\cal L}_0$
consist of elements $u\in {\cal L}_0$ such that
 $\varphi_j(u)=0$ for all
$j= 1,\ldots,p$.  Clearly,
$ \dim{\cal L}\geq k -p$.
It follows from (\ref{eq:abs0}) that $a[u,u]\leq (Au,u)$ for $u\in{\cal L}$. Therefore
(\ref{eq:pr4a}) implies (\ref{eq:abs3}). So, by Lemma~\ref{ab1}, the negative spectrum of  $H$
contains at least $k -p$ eigenvalues. It remains to take into account that $k$ is arbitrary.
$\quad\Box$

{\bf 3.}
To calculate the number of  negative eigenvalues of the operator $H$, we need several auxiliary
assertions. The first of them has a purely algebraic nature and is quite standard.

\begin{lemma}\label{alg}
 Let ${\goth R}$ be some linear space $($perhaps of infinite dimension$)$ and let $\varphi_1,\ldots,
\varphi_n,\varphi:{\goth R}\rightarrow{\Bbb C}$ be linear functionals. Suppose that  
$\varphi_1,\ldots,
\varphi_n$ are linearly independent and  denote by  ${\cal N}$  the set of their common zeros, i.e.
$u\in {\cal N}$ iff $\varphi_j(u)=0$ for any
$j= 1,\ldots,n$. Assume that $\varphi(u)=0$ for all $u\in {\cal N}$. Then $\varphi=\sum_{j=1}^n
\alpha_j\varphi_j$ for some $\alpha_j\in {\Bbb C}$.
\end{lemma}

To exclude the case when the sums in (\ref{eq:abs1}) are degenerate, we introduce the
following  

\begin{definition}\label{SI}
 Let ${\goth R}\subset {\cal H}$ be a linear set dense in ${\cal H}$ and let
$\varphi_1,\ldots,\varphi_n$ be linear functionals  defined on ${\goth R}$. We call
$\varphi_1,\ldots,\varphi_n$  strongly linear  independent if the inequality 
\begin{equation} 
|\sum_{j=1}^n \alpha_j \varphi_j(u)|\leq C ||u||,
\quad \forall u\in{\goth R},
\label{eq:SLI}\end{equation}
$($here $C$ is some positive constant$)$ 
implies that $\alpha_j=0$ for all $j=1,\ldots,n$.
\end{definition}

Thus, it is impossible to find a linear combination of strongly linear  independent functionals
which is a bounded functional on ${\cal H}$. Of course, the strong linear  independence ensures the
usual linear  independence, and every functional from a strongly linear  independent system is
unbounded.

\begin{lemma}\label{unb}
 If a functional $\varphi$ $($defined on a dense set ${\goth R})$ is unbounded, then the set of its
zeros is dense.
\end{lemma}
 {\it Proof.} --
There exists a sequence $w_k\in{\goth R}$ such that $\varphi(w_k)=1$ and $||w_k||\rightarrow 0$ as
$k\rightarrow\infty$. Moreover, for any given sequence $\eta_k\rightarrow 0$, choosing a
subsequence of $w_k$, we can satisfy the bound $||w_k|| \leq \eta_k$. Let $h\in{\cal H}$ be an
arbitrary element. Then there exists a sequence $u_k \in{\goth R}$ such that $u_k\rightarrow h$ in
${\cal H}$ as $k\rightarrow\infty$. Put
\[
u_k^{(0)}=u_k - \varphi(u_k) w_k.
\]
Clearly, $\varphi(u_k^{(0)})=0$ and $u_k^{(0)}\rightarrow h$  provided $|\varphi(u_k)|\: ||
w_k||\rightarrow 0$ as $k\rightarrow\infty$.
$\quad\Box$

\begin{corollary}\label{unb1}
The set of common zeros of any finite system of unbounded functionals $($in particular, of a
strongly   independent system$)$ is dense in ${\cal H}$.
\end{corollary}

Let us discuss some properties of functionals satisfying  Definition~\ref{SI}.

\begin{lemma}\label{unb2}
Let $\varphi_1,\ldots,\varphi_n$ be strongly linear  independent functionals and let ${\cal N}_m$ be
 the set of  common  zeros of functionals $\varphi_j,j=1,\ldots,n,\: j\neq m$. Then the restriction
of $\varphi_m$ on  ${\cal N}_m$ is unbounded.
\end{lemma}
 {\it Proof.} --
In the opposite case, there exists a bounded functional $\varphi$ such that 
$\varphi_m(u)=\varphi(u)$ for $u\in{\cal N}_m$. So it follows from Lemma~\ref{alg} that
$\varphi(u)=\varphi_m(u)+\sum_{j\neq m}\alpha_j \varphi_j(u)$ for all $u\in{\goth R}$. This
contradicts the strong  linear  independence of $\varphi_1,\ldots,\varphi_n$.
$\quad\Box$

\begin{lemma}\label{ab4}
 Let $\varphi_1,\ldots,\varphi_n$ be strongly linear  independent functionals   and let ${\cal L}_0$
be  a finite-dimensional subspace of ${\goth R}$. Then for any $\lambda >0$ and any $m=1,\ldots,n$
there exists a vector
$u^{(m)},\: ||u^{(m)}||=1$, such that $u^{(m)}\not\in {\cal L}_0$, $ \varphi_j(u^{(m)})=0$ for
$j=1,\ldots,n,\: j\neq m$, and $|\varphi_m(u^{(m)})|\geq \lambda$.
\end{lemma} {\it Proof.} --
 By Lemma~\ref{unb2}, there exists a sequence $z_k$ such that
$||z_k||=1 $, $|\varphi_m(z_k)|\rightarrow\infty $ and $ \varphi_j(z_k)=0$ for all
$j=1,\ldots,n,\: j\neq m$. Since $\dim{\cal L}_0 <\infty$, we have that
\[
\sup_{z\in {\cal L}_0,\,||z||=1} |\varphi_m(z)|<\infty.
\]
Therefore $z_k\not\in {\cal L}_0$ and $|\varphi_m(z_k)|\geq \lambda$ for  $k$
large enough. So we can set $u^{(m)}=z_k$ for such $k$. $\quad\Box$

This assertion can be generalized.

\begin{lemma}\label{ab5}
Under assumptions of Lemma~$\ref{ab4}$, for any $\lambda>0$ and any $m\leq n$,
 there exists a set of normalized vectors $u_i\not\in {\cal L}_0$, $i=1,\ldots,m$, such that
\begin{equation}
 \varphi_j(u_i)=0,\quad  j=1,\ldots, n,\; j\neq i,
\label{eq:pr1}\end{equation}
 and
\begin{equation}
 |\varphi_i(u_i)|\geq \lambda.
\label{eq:pr2}\end{equation}
 Moreover, for the subspace ${\cal L}\subset{\goth R}$  spanned by ${\cal L}_0$ and $u_1,\ldots,u_m$
\begin{equation}
\dim {\cal L} = \dim {\cal L}_0+m. 
\label{eq:pr3}\end{equation}
\end{lemma}
 {\it Proof.} --
In the case $m=1$  Lemmas~\ref{ab4} and \ref{ab5} councide. Suppose that we have already constructed
$u_1,\ldots, u_{m-1}$ such that relations (\ref{eq:pr1}), (\ref{eq:pr2}) are satisfied for
$i=1,\ldots, m-1$ and  the subspace $\tilde{\cal L}$  spanned by ${\cal L}_0$ and
$u_1,\ldots,u_{m-1}$ has dimension $\dim {\cal L}_0+m-1$. Then  existence of the vector $u_m$
with all necessary properties follows again from Lemma~\ref{ab4} applied to the subspace $\tilde{\cal
L}$. $\quad\Box$

{\bf 4.}
In Definition~\ref{SI}, ${\goth R}$ is an arbitrary linear set but, if  ${\goth R}$ is a Banach 
space, we suppose that the functionals $\varphi_1,\ldots, \varphi_n$ are bounded on this space (but  
not on ${\cal H}$, of course).  In our study of the negative spectrum of the operator $H$, the role
of  ${\goth
R}$ is played by the space ${\cal R}=H_0^{1/2} {\cal D}(H_0^{1/2})$ endowed with the norm
(\ref{eq:R}). Now we are in position to formulate the main result of this section.

\begin{theorem}\label{abs2}
 Let  $\varphi_1,\ldots,\varphi_n$ be strongly linear  independent functionals defined on ${\cal
R}$ and let  $A$ be a bounded self-adjoint operator.  Assume that equality
$(\ref{eq:abs1})$ holds
for all $u\in {\cal R}$. Then numbers $(\ref{eq:N})$ and $(\ref{eq:M})$ are related  by the equality 
\begin{equation}
 N=M+m.
\label{eq:abs2}\end{equation}
\end{theorem}
{\it Proof.} --
 First, we check that $N\leq M+m$. By  Lemma~\ref{ab1}, there exists a subspace ${\cal L}\subset
{\cal R}$ such that (\ref{eq:abs3}) is fulfilled and  $\dim {\cal L}=N$ ($\dim {\cal L}$ is an
arbitrary large number if $N=\infty$). It follows from (\ref{eq:abs1}) that  (\ref{eq:pr4a}) is
satisfied for
\[ 
{\cal L}_0=\{u\in {\cal L}: \varphi_j(u)=0,\; j=1,\ldots, m\}.
\] 
Clearly, $\dim {\cal L}_0 \geq \dim {\cal L}-m$.
Since $ M\geq \dim{\cal L}_0$, this implies that $ M\geq N-m$. 

 It remains to check that $N\geq
M+m $. Let the set ${\cal N}\subset {\cal R}$ be defined by the condition:
$u\in{\cal N}$ iff
$\varphi_j(u)=0$ for all $j=1,\ldots,n$. By Corollary~\ref{unb1}, this set  is dense in 
${\cal H}$. Therefore, by Lemma~\ref{ab2},  $M$ equals the maximal dimension of ${\cal
L}_0\subset{\cal N}$ where (\ref{eq:pr4a}) holds. Let ${\cal L}$ be the subspace   constructed in
Lemma~\ref{ab5} for ${\goth R}={\cal R}$ and sufficiently large $\lambda$ which will be chosen
later.  According to (\ref{eq:pr3}), we need only to verify relation (\ref{eq:abs3}) on this
subspace.  Every vector
$u\in{\cal L}$ has the form 
\begin{equation}
u=u_0+\sum_{i=1}^m \beta_i u_i,\quad {\rm where}\quad u_0\in{\cal L}_0,\quad \beta_j\in{\Bbb C},
\label{eq:p1}\end{equation}
so that, for $B=A+I$,
\begin{equation} 
(B u,u)= (B u_0,u_0)+ 2\Re \sum_{i=1}^m \bar{\beta}_i(B u_0, u_i) +  \sum_{i,j=1}^m
\beta_i\bar{\beta}_j(B u_i, u_j).
\label{eq:p4}\end{equation}
Recall that ${\cal L}_0\subset {\cal N}$ and, consequently, $\varphi_j(u_0 )=0$ for all
$j=1,\ldots, n$. Therefore it follows from (\ref{eq:pr1}) and (\ref{eq:p1}) that 
\begin{equation}
 \varphi_i(u )=\beta_i \varphi_i(u_i),\quad i=1,\ldots, m,
\quad {\rm and} \quad
 \varphi_j(u )=0,\quad j=m+1,\ldots, n.
\label{eq:p3}\end{equation}
Comparing (\ref{eq:p4}) and (\ref{eq:p3}), we obtain the   bound for the  form
(\ref{eq:abs1}):
\begin{eqnarray*}
 a[u,u] + || u ||^2 = (Bu,u)  -\sum_{i=1}^m   |\varphi_i(u)|^2
\\
\leq (Bu_0,u_0) + 2|| Bu_0|| \sum_{i=1}^m |\beta_i| +m b \sum_{i=1}^m
|\beta_i|^2 -\sum_{i=1}^m |\beta_i|^2 |\varphi_i(u_i)|^2,
\end{eqnarray*}
where $b=\max_{1\leq i,j\leq m}|(B u_i, u_j)|$.

Thus, for the proof of (\ref{eq:abs3}), it suffices to check that for all $u_0\in{\cal L}_0$ and any
numbers $\beta_i$
\begin{equation}
(m b - |\varphi_i(u_i)|^2)|\beta_i|^2 + 2|| Bu_0|| \: |\beta_i| + m^{-1} 
 (Bu_0,u_0) < 0,\quad i=1,\ldots,m.
\label{eq:p6}\end{equation}
Since $ (Bu_0,u_0) < 0$ and the dimension of ${\cal L}_0$ is finite,  we have that
\[
 (Bu_0,u_0)  \leq -b_0 || u_0||^2 \quad {\rm and}\quad || B u_0||\leq b_1 || u_0||
\] 
for some $b_0, b_1>0$. Taking also into account (\ref{eq:pr2}), we see that (\ref{eq:p6}) is
satisfied if
\[
(\lambda^2 -m b)|\beta_i|^2 - 2b_1   || u_0||\: |\beta_i| + m^{-1} b_0 || u_0||^2 >0.
\]
The last inequality holds for arbitrary $\beta_i$ if
$\lambda$ is large enough, that is   $\lambda^2> m  (b+b_1^2 b_0^{-1})$.
 This concludes the proofs of (\ref{eq:abs3}) and hence of Theorem~\ref{abs2}.$\quad\Box$

{\bf 5.}
It is easy to extend Proposition~\ref{BSs} and Theorem~\ref{abs2} to the case of unbounded
operators $A$. We formulate such results but do not use them in the sequel.

\noindent{\bf Proposition~\ref{BSs} bis} 
{\it Let $A$ be a self-adjoint operator in the space ${\cal
H}$ and let ${\cal R}_0 \subset {\cal R}\cap {\cal D}(A)$ be  a linear set dense in ${\cal R}$ in
the ${\cal R}$-metrics and  dense in ${\cal D}(A)$ in the $A$-metrics. 
If equality $(\ref{eq:aA})$ holds for all $u\in{\cal R}_0$, then $N=M$}.

\noindent{\bf Theorem~\ref{abs2}  bis}  {\it Let $A$ be a self-adjoint operator in the space ${\cal
H}$ and let ${\cal R}_0 \subset {\cal R}\cap {\cal D}(A)$ be  a linear set dense in ${\cal R}$ in the
${\cal R}$-metrics and  dense in ${\cal D}(A)$ in the $A$-metrics. Assume that functionals
$\varphi_1,\ldots, \varphi_n$ are strongly linear independent in the Hilbert space ${\cal D}(A)$
$($that is $||u||$ in the right-hand of $(\ref{eq:SLI})$ is replaced by $||u||_A$ $)$. Then
equality $(\ref{eq:abs2})$ is fulfilled}.

Proofs of  Proposition~\ref{BSs} bis and Theorem~\ref{abs2} bis are practically the same as those
of  Proposition~\ref{BSs} and Theorem~\ref{abs2}.

\section {the Friedrichs model}

{\bf 1.}
Our study of the discrete spectrum in the Friedrichs model relies on
the Mellin transform ${\bf M}$ defined by the equality
\begin{equation}
 ({\bf M} u) (\lambda)= (2\pi)^{-1/2}  \int_0^\infty x^{-1/2 -i\lambda} u(x)dx.
\label{eq:v8}\end{equation} 
The operator ${\bf M}:L_2({\Bbb R}_+)\rightarrow L_2({\Bbb R})$ is
unitary.

\begin{lemma}\label{M}
  Suppose that a function $b(t)$ is locally  bounded on  $(0,\infty)$ and the integral
\[
 \int_0^\infty b(t) t^{-1/2 -i\lambda} dt=:\beta(\lambda)
\]
converges at $t=0$ and $t=\infty$ uniformly in  $\lambda\in {\Bbb R}$.  Then for any function
$u\in C_0^\infty ({\Bbb R}_+)$
\begin{equation}
\int_0^\infty \int_0^\infty b  (xy)  u(y) \overline{u(x)}dx  dy=\int_{-\infty}^\infty \beta
(\lambda)  ({\bf M}u)(-\lambda) \overline{({\bf M}u)(\lambda)}d\lambda.
\label{eq:v10}\end{equation}
\end{lemma}
 {\it Proof.} -- 
 Changing in the left-hand side the variables $x=e^t$, $y=e^s$ and denoting
$u_1(t)= e^{t/2} u(e^t)$, $b_1(t)= e^{t/2} b(e^t)$, we rewrite it as 
\[
\int_{-\infty}^\infty \int_{-\infty}^\infty b_1  (t+s)  u_1(s) \overline{u_1(t)}dt ds.
\]
 If $u_1(t)=0$ for $|t|\geq n$, then, by virtue of the convolution formula and the Parseval 
identity, this integral equals
\begin{equation}
\int_{-\infty}^\infty \beta_n (\lambda)  {\hat u}_1(-\lambda) \overline{{\hat u}_1(\lambda)}d\lambda,
\label{eq:v10a}\end{equation} 
where ${\hat u}_1$ is the Fourier transform of $u_1$ and 
\[
 \beta_n (\lambda) =\int_{-2n}^{2n} b_1(t) e^{-i\lambda t} dt.
\] 
 Under our assumptions  functions $\beta_n (\lambda)$ are uniformly bounded and converge to
$\beta(\lambda)$ as
$n\rightarrow\infty$. Since ${\hat u}_1={\bf M}u$ belongs to the Schwartz space ${\cal S}({\Bbb R})$,
we can pass to the limit $n\rightarrow\infty$ in integral (\ref{eq:v10a}).  The expression obtained
equals the right-hand side of (\ref{eq:v10}). $\quad\Box$

Under assumptions of Lemma~\ref{M}, the function $\beta(\lambda)$ is of course continuous and
bounded.

Let us define a unitary mapping $U:L_2({\Bbb R})\rightarrow L_2 ({\Bbb R}_+;{\Bbb C}^2)$ and a
$2\times 2$- matrix ${\cal B}(\lambda)$ by the equalities
\begin{equation}
 (U w)(\lambda)=\Bigl(\begin{array}{c}w(\lambda)\\w(-\lambda)\end{array}\Bigr),\quad {\cal
B}(\lambda)=\Bigl(\begin{array}{cc} 0& \beta (\lambda)\\
\overline{\beta (\lambda)}& 0\end{array}\Bigr),\quad \lambda>0.
\label{eq:v11}\end{equation}
If $\beta(-\lambda)= \overline{\beta (\lambda)}$, then
\begin{equation}
 \int_{-\infty}^\infty \beta (\lambda)  w(-\lambda) \overline{w(\lambda)}d\lambda=({\cal B} U w, U
w)_{L_2 ({\Bbb R}_+;{\Bbb C}^2)},
\label{eq:v12}\end{equation}
  where ${\cal B}$ is the operator of multiplication by ${\cal B} (\lambda)$.
Since eigenvalues of the matrix ${\cal B}(\lambda)$ equal $\pm |\beta(\lambda)|$, we have

\begin{lemma}\label{sp} 
Let $\beta (\lambda)$ be a continuous function of $\lambda\in (0,\infty)$ and
\[
 p =\max_{\lambda\in{\Bbb R}_+} |\beta (\lambda)|,\quad q =\min_{\lambda\in{\Bbb R}_+}
|\beta(\lambda)|
\]
$($the case $p=\infty$ is not excluded$)$. Then
 the spectrum of ${\cal B}$ consists of the union $[-p,-q]\cup [q ,p]$.
\end{lemma}

{\bf 2.}
Let now  ${\cal H}=L_2({\Bbb R}_+)$, let $H_0$ be multiplication by the function $x^{2l},\; l>0,$
 and let an integral  operator $V$ be defined by formula (\ref{eq:v1}). We suppose that the  
function $v(t)=\overline{v(t)}$ is locally bounded on $(0,\infty)$. Our assumption on its
behaviour as $t\rightarrow\infty$    will be made in terms of the Mellin transform.

\begin{assumption}\label{ass}
The    integral 
\[
  \int_1^\infty v (t) t^{-1/2-l-i\lambda} dt
\]
 converges uniformly $($but perhaps not absolutely$)$        in $\lambda\in{\Bbb R}$.
\end{assumption}

The precise definition of the operator $H_\gamma=H_0+ \gamma V$ where a coupling constant
$\gamma\in{\Bbb R}$ can be given on the basis of the following

\begin{lemma}\label{comp} 
Let Assumption~$\ref{ass}$ hold and let $v(t)=O(t^r)$ with $r>-1/2$ as $t\rightarrow 0$. Then the
operator
$T=(H_0+I)^{-1/2}  V (H_0+I)^{-1/2}$ is compact.
\end{lemma}
{\it Proof.} -- 
Let $\chi_R $ be  the characteristic function of the interval $(0,R)$ and $\tilde{\chi}_R=1-\chi_R
$. Denote by
$V_R$  and $\tilde{V}_R$  the integral operators with kernels $v(xy)\chi_R (xy)$ and
$v(xy)\tilde{\chi}_R (xy)$, respectively. First, we check that the operator
$(H_0+I)^{-1/2}  V_R (H_0+I)^{-1/2}$ is compact for any $R>0$. Indeed, let us consider the
operator-function
\[
F(z)= (H_0+I)^{-z}  V_R (H_0+I)^{-z},\quad \Re z\geq 0.
\]
The function $v(t)\chi_R (t)$ satisfies the assumptions of
 Lemma~\ref{M} so that  $V_R$ is a bounded operator. Consequently, the operators $F(z)$ are bounded
for all
$\Re z\geq 0$, and the function $F(z)$ is analytic for $\Re z > 0$ and is continuous in $z$ up to the
line
$\Re z = 0$. On the other hand, $F(z)$  is the integral operator with kernel 
\[
(x^{2l}+1)^{-z}v(xy)\chi_R (xy)(y^{2l}+1)^{-z}.
\]
So $F(z)$ belongs to the Hilbert-Schmidt class if $4l\Re z>1$. By complex interpolation,
this implies that $F(z)$ is compact for all $\Re z > 0$.

To finish the proof, it suffices to show that the norm of  the operator 
$B_R=H_0^{-1/2}\tilde{V}_R   H_0^{-1/2}$ tends to zero as
$R\rightarrow\infty$. Clearly,
  $B_R$ is also the integral operator with kernel
$b_R(xy)=(xy)^{-l}v(xy)\tilde{\chi}_R (xy)$. By virtue of Assumption~\ref{ass}, we can apply
Lemmas~\ref{M} and \ref{sp} to it. This implies that  
\[
||B_R||= \max_{\lambda} \left|\int_R^\infty v(t) t^{-1/2-l-i\lambda}dt\right| \rightarrow 0 \quad
{\rm as} \quad R\rightarrow\infty.\quad\Box
\]

Thus, equality (\ref{eq:BS0}) holds with a compact operator $T$ and hence the operator
$H_\gamma=H_0+ \gamma V$ can be defined as a self-adjoint operator  in
terms of the corresponding quadratic form. Moreover, $H_\gamma$ is semi-bounded from below  and
$\sigma_{ess}(H_\gamma)=[0,\infty)$ for any  $\gamma\in{\Bbb R}$.

To study the discrete spectrum of $H_\gamma$, we need additional conditions on $v(t)$ as
$t\rightarrow 0$.

\begin{assumption}\label{ass1}
Suppose that, as $t\rightarrow 0$,
\begin{equation}
v(t)=\sum_{k=1}^N v_k t^{r_k} + O(t^{r_{N+1}}),
\label{eq:as1}\end{equation} 
where $-1/2<r_1<\cdots <r_N < r_{N+1}$ and $l< r_{N+1} +1/2$.
\end{assumption}

We assume that $v_k\neq 0$ for $k=1,\ldots, N$ but do not exclude the case when the sum in
(\ref{eq:as1}) is absent, that is
\begin{equation}
 v(t)=O(t^{r_1})\quad {\rm with}\quad r_1>l-1/2.
\label{eq:as2}\end{equation}
 We set
\begin{equation}
 b_l (t)=t^{-l} (v(t)-\sum_{r_k <l-1/2}v_k t^{r_k})  
\label{eq:w2}\end{equation}
($b_l (t)=t^{-l}v (t)$ in the case (\ref{eq:as2}))
and
\begin{equation}
\beta_l^{(\kappa)}(\lambda)=\int_0^\infty 
\chi^{(\kappa)}(t) b_l(t) t^{-1/2-i\lambda} dt,\quad \kappa=0,1,
\label{eq:bb}\end{equation}
where  $\chi^{(0)}(t)$ and $\chi^{(1)}(t)$ are the characteristic functions of the intervals $(0,1)$ 
and $(1,\infty)$, respectively. In definition (\ref{eq:bb}) of the function $\beta_l^{(0)}$ we
suppose  that
 $l\neq r_n+1/2$ for $n=1,\ldots, N$. 

 The following assertions are quite elementary.

\begin{lemma}\label{w}
Let Assumption~$\ref{ass}$  be satisfied.
Then  the integral $(\ref{eq:bb})$ for $\kappa=1$ converges at  $t=\infty$ uniformly in
$\lambda\in{\Bbb R}$. 
\end{lemma}

\begin{lemma}\label{wa}
 Let Assumption~$\ref{ass1}$ be satisfied.  Suppose that
 $l\neq r_n+1/2$ for $n=1,\ldots, N$.  Then  the integral
$(\ref{eq:bb})$ for $\kappa=0$
 converges at $t=0$ uniformly in $\lambda\in{\Bbb R}$.
\end{lemma}

If conditions of both Lemmas~\ref{w} and \ref{wa} are fulfilled, then the function
\begin{equation}
\beta_l(\lambda)=\int_0^\infty b_l(t) t^{-1/2 -i\lambda} dt
\label{eq:be}\end{equation}
 is continuous and
\[
 p_l=\max_{\lambda\in{\Bbb R}}|\beta_l(\lambda)|<\infty.
\]

Let us denote by $N_l^{(\pm)}$ the number of  $k= 1,\ldots,N$ such that $r_k<l-1/2$ and
$\mp v_k>0$. In the case (\ref{eq:as2}) we set $N_l^{(\pm)}=0$.
The main result of this section is formulated in the following

\begin{theorem}\label{Fr}
 Let Assumptions~$\ref{ass}$ and $\ref{ass1}$ be satisfied.   

$1^0$   Suppose that
 $l\neq r_n+1/2$ for $n=1,\ldots, N$. Put
$\sigma_l=p_l^{-1}$. Then the negative spectrum of the operator $H_\gamma$ is infinite if 
$|\gamma|>\sigma_l$. In the case $\pm\gamma\in (0,\sigma_l]$,   it consists of $N_l^{(\pm)}$
eigenvalues.

$2^0$ If $l= r_n+1/2$ for some $n=1,\ldots, N$, then the negative spectrum of the operator
$H_\gamma$ is infinite for any $\gamma\neq 0$.
\end{theorem}

We start the proof with calculating the quadratic form (\ref{eq:BS1}), which equals now
\begin{equation}
 a_l[u,u]=\gamma\int_0^\infty \Bigl(\int_0^\infty v (xy) y^{-l} u(y) dy\Bigr) x^{-l}
\overline{u(x)}dx.
\label{eq:v3}\end{equation}
Using notation (\ref{eq:w2}), we can rewrite (\ref{eq:v3}) for $u\in C_0^\infty ({\Bbb R}_+)$ as 
\begin{equation}
 a_l[u,u]=\gamma  b_l[u,u] + \gamma\sum_{r_k <l-1/2}  v_k |\Phi_{r_k-l}(u)|^2,
\label{eq:v4}\end{equation}
where
\[
 b_l[u,u]=\int_0^\infty \int_0^\infty b_l (xy)  u(y) \overline{u(x)}dx  dy,
\]
and
\begin{equation}
 \Phi_p(u)=   \int_0^\infty x^p u(x)dx.
\label{eq:v7}\end{equation}
Of course, in the case (\ref{eq:as2}) the sum in (\ref{eq:v4}) is absent.
By definition (\ref{eq:R}), the set ${\cal R}={\cal R}^{(l)}$ consists now of functions $u$ such
that 
\[
 ||u||^2_{\cal R} = \int_0^\infty (1+x^{-2l})|u(x)|^2dx <\infty.
\]
It follows from Lemma~\ref{comp} that  the form (\ref{eq:v3})  is bounded on ${\cal R}^{(l)}$.

\begin{lemma}\label{ind}
Let $p_k\in (-1/2-l,-1/2)$ for $k=1,\ldots,n$. Then
  the functionals $\Phi_{p_k}(u)$  defined by
$(\ref{eq:v7})$   are bounded on ${\cal R}^{(l)}$ and are strongly linear
independent.
\end{lemma}
 {\it Proof.} --
The inequality 
\[ 
|\Phi_{p} (u)| \leq C ||u||_{\cal R}
\]
 is equivalent to the inclusion
\[
 x^{p} (1+x^{-2l})^{-1/2}\in L_2 ({\Bbb R}_+)
\]
 which is true if $p\in (-1/2-l,-1/2)$.
The functionals $\Phi_{p_1},\ldots, \Phi_{p_n}$ are strongly linear independent because
a function $\sum_{k=1}^n c_k
x^{p_k}$ does not belong
to the space
$L_2({\Bbb R}_+)$ unless all $c_k=0.\quad\Box$

Put $b_l^{(\kappa)}(t)=\chi^{(\kappa)}(t) b_l(t)$.
 With the help of  Lemmas~\ref{M}, \ref{w} and \ref{wa} it easy to show that
\begin{equation}
\int_0^\infty \int_0^\infty b_l^{(\kappa)}  (xy)  u(y) \overline{u(x)}dx  dy=\int_{-\infty}^\infty
\beta_l^{(\kappa)} (\lambda)  ({\bf M}u)(-\lambda) \overline{({\bf M}u)(\lambda)}d\lambda.
\label{eq:v10x}\end{equation}
 The precise statements are formulated in the two following assertions.

\begin{lemma}\label{inf}
 Let Assumption~$\ref{ass}$ hold and $u\in C_0^\infty ({\Bbb R}_+)$. Then  representation
$(\ref{eq:v10x})$ is valid for $\kappa=1$.
\end{lemma}

\begin{lemma}\label{inf1}
 Let Assumption~$\ref{ass1}$ hold and $u\in C_0^\infty ({\Bbb R}_+)$.
 Suppose that $l\neq r_n+1/2$ for $n=1,\ldots, N$.  Then representation $(\ref{eq:v10x})$ is valid
for $\kappa=0$.
\end{lemma}

To check the part $1^0$ of Theorem~\ref{Fr}, we compare Lemmas~\ref{inf}, \ref{inf1}  and take into
account equality (\ref{eq:v12}). This yields the representation
\begin{equation}
 b_l[u,u]=({\cal B}_l U {\bf M} u,  U {\bf M} u)_{L_2 ({\Bbb R}_+;{\Bbb C}^2)},
\label{eq:v9a}\end{equation} 
where ${\cal B}_l$ is multiplication by matrix (\ref{eq:v11}) with elements (\ref{eq:be}). It
follows from (\ref{eq:v4}) and (\ref{eq:v9a}) that for all $u\in C_0^\infty({\Bbb R}_+)$
\begin{equation}
 a_l[u,u]=(A_l u,u) + \gamma\sum_{r_k<l-1/2}v_k |\Phi_{r_k-l}(u)|^2,
\label{eq:ZZ1}\end{equation}
where
\begin{equation}
 A_l =\gamma (U {\bf M})^\ast {\cal B}_l U {\bf M}
\label{eq:ZZ2}\end{equation}
is a bounded operator in ${\cal H}$. Since, by Lemma~\ref{ind}, the functionals $\Phi_{r_k-l}(u)$ are
bounded on ${\cal R}^{(l)}$, equality (\ref{eq:ZZ1}) extends by continuity to all $u\in {\cal
R}^{(l)}$. This gives us representation (\ref{eq:abs1}) where $m=N_l^{(+)}$, $n-m=N_l^{(-)}$
if $\gamma>0$ and $m=N_l^{(-)}$, $n-m=N_l^{(+)}$ if $\gamma <0$.  According to Lemma~\ref{ind} the
corresponding functionals $\varphi_1,\ldots, \varphi_n$ are strongly linear independent.
By virtue of Lemma~\ref{sp}, the total multiplicity of the spectrum of  operator (\ref{eq:ZZ2}) in
the interval $(-\infty,-1)$ is zero if $|\gamma| p_l \leq 1$ and it is infinite if 
$|\gamma| p_l > 1$.  Thus the assertion of the part $1^0$ of Theorem~\ref{Fr} follows immediately
from Theorem~\ref{abs2}.

To check the part $2^0$ of Theorem~\ref{Fr}, we verify the assumptions of
Theorems~\ref{abs1}. We rely again on 
representation (\ref{eq:v4}) valid at least for $u\in C_0^\infty ({\Bbb R}_+)$. 
 Let ${\cal R}_0$
consist of functions $u\in C_0^\infty ({\Bbb R}_+)$ such that 
\begin{equation}
\Phi_{-1/2}(u)=\int_0^\infty u(x) x^{-1/2} dx=0.
\label{eq:s1}\end{equation} 
 Let us extend representation (\ref{eq:v10x}) for the form $b_l^{(0)}[u,u]$  
 to the case $l=r_n+1/2$.

\begin{lemma}\label{inf2}
 Let Assumption~$\ref{ass1}$ hold and $u\in {\cal R}_0$. Then representation
$(\ref{eq:v10x})$ is valid for $\kappa=0$ and  $l=r_n+1/2$ with the function
\begin{equation}
 \beta_{r_n+1/2}^{(0)} (\lambda) = \int_0^1 \Bigl(v (t)-\sum_{k=1}^n
 v_k t^{r_k}\Bigr) t^{-1-r_n-i\lambda} dt + i v_n \lambda^{-1}.
\label{eq:d1}\end{equation}
\end{lemma}
 {\it Proof.} -- 
Let us proceed from equality (\ref{eq:v10x}) for $\kappa=0$ and $l=r_n+1/2-\varepsilon$,
$\varepsilon>0$. Then  we pass to the limit $\varepsilon\rightarrow 0$ in this equality. Its
left-hand side is of course continuous in $l$ for any $u\in C_0^\infty ({\Bbb R}_+)$. Let us verify
the convergence of the right-hand side as $\varepsilon\rightarrow 0$. 
Comparing definition (\ref{eq:w2}), (\ref{eq:bb}) of the functions $\beta_l^{(0)} (\lambda)$ with
(\ref{eq:d1}), we see that
\[
 \beta_l^{(0)} (\lambda) = \int_0^1 \Bigl(v (t)-\sum_{k=1}^n
 v_k t^{r_k}\Bigr) t^{-1/2 -l -i\lambda} dt + v_n (r_n+1/2-l-i\lambda)^{-1}.
\]
These functions converge as $\varepsilon\rightarrow 0$ to function (\ref{eq:d1}) uniformly
in $\lambda$ outside of any neighbourhood of the point $\lambda=0$ and
\[
 |\beta_l^{(0)} (\lambda)|\leq C (|\lambda|^{-1}+1).
\] 
 This suffices to justify passing to the limit in the integral
over $\lambda$ because ${\bf M}u\in{\cal S}({\Bbb R})$ and $({\bf
M}u)(0)=0$ by virtue of condition (\ref{eq:s1}). $\quad\Box$

Let now $l=r_n+1/2$, 
$\beta_l(\lambda)=\beta_l^{(0)}(\lambda)+\beta_l^{(1)}(\lambda)$
 where  $\beta_l^{(0)}$ and $\beta_l^{(1)}$ are given by
(\ref{eq:d1}) and (\ref{eq:bb}), respectively, and let ${\cal B}_l$ be multiplication by
matrix (\ref{eq:v11}) with the elements
 $\beta_l(\lambda)$. Lemmas~\ref{inf} and \ref{inf2} imply that, in the case $l=r_n+1/2$,
equality  (\ref{eq:v9a}) holds for $u\in {\cal R}_0$. This gives us representation (\ref{eq:ZZ1})
where $u\in {\cal R}_0$ and the operator $A_l$ is defined by equality (\ref{eq:ZZ2}).
 Since
\begin{equation}
\lim_{\lambda\rightarrow 0}|\lambda|\:|\beta_l(\lambda)|=  |v_n| \neq 0,\quad l=r_n+1/2,
\label{eq:sbe}\end{equation}
 it follows from  Lemma~\ref{sp} that the operator ${\cal B}_l$ is unbounded from below and
consequently 
\[
\dim E_{A_l}(-\infty,-1)=\infty
\]
for $l=r_n+1/2$ and any $\gamma\neq 0$. Note also the following elementary

\begin{lemma}\label{dense}
The set ${\cal R}_0$ is dense in  ${\cal D}( A_l)$ where $l=r_n+1/2$.
\end{lemma}
 {\it Proof.} -- 
Recall that the function $\beta_l(\lambda)$ is bounded except the point $\lambda=0$ where
it satisfies (\ref{eq:sbe}). Therefore it follows from (\ref{eq:v11}) and (\ref{eq:ZZ2}) that
the inclusion $u\in{\cal D}( A_l)$ is equivalent to the bound
\begin{equation}
\int_{-\infty}^\infty (1+\lambda^{-2}) |w(\lambda)|^2 d\lambda <\infty\quad 
{\rm for}\quad w={\bf M}u.
\label{eq:dn}\end{equation}
Clearly, the Mellin transform (\ref{eq:v8}) can be factorized as ${\bf M}=\Phi G$ where $\Phi$ is
the Fourier transform in $L_2({\Bbb R})$ and
$(Gu)(t)=e^{t/2}u(e^t)$. In terms of $g(\lambda)=\lambda^{-1} w(\lambda)$, (\ref{eq:dn})  is
equivalent to the condition
\begin{equation}
\int_{-\infty}^\infty (|\tilde{g}(t)|^2 + |\tilde{g}^\prime (t)|^2)dt <\infty,
\quad \tilde{g}=\Phi^\ast g. 
\label{eq:dn1}\end{equation}
Of course, there exists a sequence $\tilde{g}_j \in C_0^\infty ({\Bbb R})$ such that $\tilde{g}_j$
converge to $\tilde{g}$ as $j\rightarrow\infty$ in the metrics (\ref{eq:dn1}) 
(that is in the Sobolev space $H^1 ({\Bbb R})$). Set $\tilde{w}_j (t) =-i\tilde{g}_j^\prime (t)$ so
that $w_j (\lambda) =\lambda g_j(\lambda)$. Then $w_j$ converge to $w$ in the metrics
(\ref{eq:dn}). Moreover,  $\tilde{w}_j \in C_0^\infty ({\Bbb R})$ and
\[
\int_{-\infty}^\infty \tilde{w}_j (t) dt=-i \int_{-\infty}^\infty \tilde{g}_j^\prime (t) dt=0.
\]
It follows that $u_j=G^\ast \tilde{w}_j\in C_0^\infty ({\Bbb R}_+)$, $u_j$ satisfy (\ref{eq:s1}) and
$u_j\rightarrow u$ in  ${\cal D}(A_l)$ as $j\rightarrow\infty.\quad\Box$

Thus, we have verified all conditions of Theorem~\ref{abs1} and hence the negative spectrum of the
operator
$H_\gamma$ is infinite for any
$\gamma\neq 0$. This concludes the proof of Theorem~\ref{Fr}.

{\bf 4.}
The function (\ref{eq:be}) can be calculated on the basis of the following

\begin{proposition}\label{mer}
 Suppose that the integral
\[
{\cal V}(T)=\int_1^T v(t) t^{-1/2} dt
\]
is bounded uniformly in $T\geq 1$ and that Assumption~$\ref{ass1}$ holds. Then the function
\begin{equation}
{\goth B}(z)=\int_0^\infty v(t) t^{-1/2-z} dt
\label{eq:mer0}\end{equation}
is analytic in the band $\Re z\in(0,r_1+1/2)$ and admits a meromorphic continuation in the band $\Re
z\in(0,r_{N+1}+1/2)$. The function ${\goth B}(z)$ has only simple poles in the points $r_n+1/2$ with
the residues $-v_n$, $n=1,\ldots,N$. Moreover, it is given by the formula
\begin{equation}
 {\goth B}(z)=\int_0^\infty\Bigl( v(t)-\sum_{k=1}^n v_k t^{r_k}\Bigr) t^{-1/2-z} dt
\label{eq:mer1}\end{equation}
in the band $\Re z\in(r_n+1/2,r_{n+1}+1/2)$.
\end{proposition}
 {\it Proof.} -- 
Integrating by parts, we see that the integral
\[
 {\goth B}_1(z)=\int_1^\infty v(t) t^{-1/2-z} dt=z \int_1^\infty {\cal V}(t) t^{-1-z} dt 
\]
defines an analytic function for all $\Re z>0$. If  $\Re z> a_n+1/2$, then
\begin{equation}
 {\goth B}_1(z)=\int_1^\infty \Bigl(v(t)-\sum_{k=1}^n v_k t^{r_k} \Bigr)t^{-1/2-z} dt
+\sum_{k=1}^n v_k (z-r_k-1/2)^{-1}.
\label{eq:mer2}\end{equation}
Similarly, if $\Re z< a_1+1/2$, then
\begin{equation}
 {\goth B}_0(z)= \int_0^1 v(t) t^{-1/2-z} dt =\int_0^1 \Bigl(v(t)-\sum_{k=1}^n v_k t^{r_k}
\Bigr)t^{-1/2-z} dt -\sum_{k=1}^n v_k (z-r_k-1/2)^{-1}
\label{eq:mer3}\end{equation}
According to Assumption~$\ref{ass1}$, the  integral in the
right-hand side of (\ref{eq:mer3}) is analytic for $\Re z< r_{n+1}+1/2$ so that (\ref{eq:mer3})
gives the meromorphic continuation of  the  function ${\goth B}_0(z)$. In particular, if $n=N$
we obtain that the function ${\goth B} (z)={\goth B}_0(z)+{\goth B}_1(z)$ is meromorphic in the band
$\Re z\in(0,r_{N+1}+1/2)$. Finally, comparing
representations (\ref{eq:mer2}) and (\ref{eq:mer3}), we arrive at (\ref{eq:mer1}). 
$\quad\Box$

Thus, to calculate the function $\beta_l(\lambda)$, it suffices to compute integral (\ref{eq:mer0})
for $\Re z\in(0,r_1+1/2)$ and then to find its meromorphic continuation into the band $\Re
z\in(0,r_{N+1}+1/2)$. Putting together relations (\ref{eq:w2}), (\ref{eq:be}) and (\ref{eq:mer1}),
we see that
\begin{equation}
\beta_l(\lambda)= {\goth B}(l+i\lambda), \quad l\neq r_n +1/2.
\label{eq:mer4}\end{equation}

\section {examples}

{\bf 1}.
Let us first consider the operator $H_\gamma=H_0+ \gamma V$ in the space $L_2({\Bbb R}_+)$.
Recall that  $H_0$ is multiplication  by $x^{2l}$ and a perturbation $V$ is defined by formula
(\ref{eq:v1}). As an example, we choose
\begin{equation}
v(t)=v_{p,q}(t)=t^q {\cal I}_p (t)
\label{eq:bes1}\end{equation}
where  ${\cal I}_p$ is the Bessel function. It follows from the asymptotics of ${\cal I}_p(t)$ at
infinity and  from its expansion at $t=0$ that
\[
 v(t)= (2/\pi)^{1/2} t^{q-1/2}\Bigl( \cos(t-(2p+1)\pi/4)+(2t)^{-1}(4^{-1}-p^{-2})
\sin(t-(2p+1)\pi/4)\Bigr)+ O(t^{q-5/2})
\]
as $ t\rightarrow\infty$ and
\begin{equation}
 v(t) = \sum_{k=0}^\infty (-1)^k 2^{-2k-p} \Bigl(k! \Gamma (k+p+1)\Bigr)^{-1} t^{p+q+2k},
\label{eq:bes3}\end{equation}
where $\Gamma(\cdot)$ is the $\Gamma$-function. Therefore the condition of Proposition~\ref{mer} on
the function ${\cal V}(T)$ is satisfied for all $q\leq 1$ and (\ref{eq:bes3}) gives us relation
(\ref{eq:as1}) with numbers
 $r_k=p+q+2k$, where $k=0,1,2,\ldots$. The corresponding coefficients $v_k$ are positive for
even
$k$ and are negative for odd $k$.  So Assumption~\ref{ass1} is fulfilled for $p+q >-1/2$ and all $l>
0$. Using formula (19), section 7.7 of  \cite{BE}, vol. 2, we find that function (\ref{eq:mer0})
equals now
\begin{equation} 
{\goth B}(z)=2^{q-z-1/2} \Gamma ((p+q-z+1/2)/2)
\Gamma^{-1} ((p-q+z +3/2)/2).
\label{eq:bes4}\end{equation}
 Let us set
\begin{equation} 
\sigma_l= 2^{-q+l+1/2}\min_{\lambda\in{\Bbb R}}| \Gamma ((p-q+l+i\lambda +3/2)/2)
\Gamma^{-1} ((p+q-l-i\lambda +1/2)/2)|.
\label{eq:bes5}\end{equation}
Remark that, by the Stirling formula, ${\goth B}(l+i\lambda) \rightarrow 0$ as
$|\lambda|\rightarrow\infty$ so that  the spectrum of the corresponding operator ${\cal B}_l$ (see
Lemma~\ref{sp}) consists of the interval $[-\sigma_l^{-1},\sigma_l^{-1}]$. In our particular case,
Theorem~\ref{Fr} gives the following assertion.

\begin{proposition}\label{BES}
Let the function $v(t)$ be given by formula $(\ref{eq:bes1})$ where
$-1/2-p <q \leq 1$.
Then the negative spectrum of the operator
$H_\gamma$ is infinite for any $\gamma\neq 0$ if $l=  p+ q+1/2+ 2k$ for some $k=0,1,2,\ldots$.
In the opposite case it is infinite if $|\gamma| > \sigma_l$ where the number $\sigma_l$ is defined
by $(\ref{eq:bes5})$. If $\gamma\in [-\sigma_l,0)$, then the negative spectrum of the operator
$H_\gamma$ is empty for $l< p+ q+1/2$ and it consists of $k+1$ eigenvalues if
$l\in(p+ q+1/2 +2k,p+ q+5/2 +2k)$. 
If $\gamma\in (0, \sigma_l]$, then the negative spectrum of the operator
$H_\gamma$ is empty for $l< p+ q+5/2$ and it consists of $k+1$ eigenvalues if
$l\in(p+ q+5/2 +2k,p+ q+9/2 +2k)$.
\end{proposition}

We note special cases $p=1/2$ when $q\in (-1,1]$ and $p=-1/2$ when $q\in (0,1]$:
\begin{equation} 
v_{ 1/2,q}(t)= (2/\pi)^{1/2} t^{q-1/2} \sin t \quad{\rm and}\quad v_{-1/2,q}(t)=
(2/\pi)^{1/2} t^{q-1/2} \cos t.
\label{eq:PC}\end{equation}

{\bf 2}.
Below we consider the operators ${\bf H}_\gamma^{(c)}={\bf H}_0+\gamma {\bf V}^{(c)}$ and ${\bf
H}_\gamma^{(s)}={\bf H}_0+\gamma {\bf V}^{(s)}$ in the space
$L_2({\Bbb R}^d)$,  where  ${\bf H}_0$ is multiplication  by $|x|^{2l}$ and ${\bf V}^{(c)}$, ${\bf
V}^{(s)}$ are defined by formulas (\ref{eq:I1}). Let ${\goth h}_n$ be the subspace of spherical
functions
$Y_n(\omega),\;
\omega\in{\Bbb S}^{d-1}$, of order $n$, let ${\cal K}$ be the $L_2$-space with weight $r^{d-1}$ of
functions defined on ${\Bbb R}_+$ and let ${\goth H}_n={\cal K}\otimes {\goth h}_n$. To put it
differently, ${\goth H}_n\subset {\Bbb R}^d$ is the subspace of functions $u_n$ of the form
\begin{equation}
u_n(x)=|x|^{-\delta} g(|x|) Y_n (\hat{x}),\quad \hat{x}=x|x|^{-1},\quad \delta=(d-1)/2,
\label{eq:ex1}\end{equation}
where $g\in L_2 ({\Bbb R}_+)$ and $Y_n \in{\goth h}_n$.  Then
\[
L_2({\Bbb R}^d)=\bigoplus_{n=0}^\infty {\goth H}_n, \quad {\goth H}_n={\cal K}\otimes {\goth h}_n,
\]
and every subspace ${\goth H}_n$ is invariant with respect to the Fourier operator $\Phi$ which
reduces to the Fourier-Bessel transform on ${\goth H}_n$. More precisely, set 
\begin{equation}
(\Phi_n g)(r)=i^{-n}\int_0^\infty (rs)^{1/2}{\cal I}_{n+(d-2)/2} (rs) g(s)ds.
\label{eq:ex11}\end{equation}
Then, for function (\ref{eq:ex1}),
\[
(\Phi u_n)(x)=|x|^{-\delta} (\Phi_n g)(|x|) Y_n (\hat{x}).
\]
The operator $\Phi_n$ is of course unitary on $L_2({\Bbb R}_+)$. It follows from (\ref{eq:ex1}) that
\[
({\bf V}^{(c)} u_n)(x)=(-1)^{n/2}|x|^{-\delta} (\Phi_n g)(|x|) Y_n (\hat{x})
\]
for  even $n$ and ${\bf V}^{(c)}  u_n=0$ for  odd $n$. Similarly,
\[
 ({\bf V}^{(s)}  u_n)(x)=(-1)^{(n+1)/2}|x|^{-\delta} (\Phi_n g)(|x|) Y_n (\hat{x})
\] 
for  odd $n$ and ${\bf V}^{(s)}  u_n=0$ for  even $n$. Let us set $\tau_n=(-1)^{n/2}$ for even
$n$,
$\tau_n=(-1)^{(n+1)/2}$ for odd $n$, 
\begin{equation}
H _\gamma^{(n)}=H_0+\tau_n\gamma  \Phi_n
\label{eq:Hn}\end{equation}
 and let $T:L_2({\Bbb
R}_+) \rightarrow{\cal K}$ be a unitary operator defined by $(Tg)(r)=r^{-\delta} g(r)$. Then 
\begin{equation}
{\bf H}_\gamma^{(c)}= \bigoplus_{m=0}^\infty T H_\gamma^{(2m)}T^\ast \otimes I_{2m},
\quad
{\bf H}_\gamma^{(s)}= \bigoplus_{m=0}^\infty T H_\gamma^{(2m+1)}T^\ast \otimes I_{2m+1},
\label{eq:DEC}\end{equation} 
where $I_n$ is the identity operator in the space ${\goth h}_n$.
Recall that
\begin{equation}
 \dim {\goth h}_n= (2n+d-2) (n+d-3)! ((d-2)! n!)^{-1}=:\nu_n.
\label{eq:ex3}\end{equation}

Comparing (\ref{eq:bes1}) and (\ref{eq:ex11}) and setting 
\begin{equation}
 p=n+(d-2)/2,\quad q =1/2,
\label{eq:SPE}\end{equation}
 we see that Proposition~\ref{BES} can be directly applied to every operator (\ref{eq:Hn}).

\begin{proposition}\label{Hn}
 The negative spectrum of the operator
$H_\gamma^{(n)}$ is infinite for any $\gamma\neq 0$ if $l=n+d/2+ 2k$ for some $k=0,1,2,\ldots$. In
the opposite case it is infinite if $|\gamma| > \sigma_l^{(n)}$ where
\begin{equation}
\sigma_l^{(n)}= 2^l\min_{\lambda\in{\Bbb R}}| \Gamma ((n+d/2+l+i\lambda )/2)
\Gamma^{-1} ((n+d/2-l-i\lambda )/2)|.
\label{eq:SPEA}\end{equation}
 If $\tau_n\gamma\in [-\sigma_l^{(n)},0)$, then the negative spectrum of the operator
$H_\gamma^{(n)}$ is empty for $l< d/2 +n$ and it consists of $k+1$ eigenvalues if
$l\in(d/2 +n +2k,d/2 +n+2 +2k)$.  If $\tau_n\gamma\in (0, \sigma_l^{(n)}]$, then the negative
spectrum of the operator
$H_\gamma^{(n)}$ is empty for $l< d/2 +n +2$ and it consists of $k+1$ eigenvalues if
$l\in(d/2 +n+ 2 +2k,d/2 +n +4 +2k)$.
\end{proposition}

Combining Proposition~\ref{Hn} with decomposition (\ref{eq:DEC}) we can deduce
results on the operators ${\bf H}_\gamma^{(c)}$ and ${\bf H}_\gamma^{(s)}$. Let us start with
exceptional values of $l$.

\begin{theorem}\label{INF}
Let $l=d/2+ 2k$ for some $k=0,1,2,\ldots$. 
Then the negative spectrum of the operator
${\bf H}_\gamma^{(c)}$ is infinite for any $\gamma\neq 0$.
Let $l=d/2+ 2k+1$ for some
$k=0,1,2,\ldots$. Then the negative spectrum of the operator
${\bf H}_\gamma^{(s)}$ is infinite for any $\gamma\neq 0$.
\end{theorem}

{\bf 3}.
To consider other values of $l$, we first find a relation between  functions
$\beta_l^{(n)}(\lambda)$ associated to different operators  $H_\gamma^{(n)}$.
 Remark that according to (\ref{eq:mer4}), (\ref{eq:bes4}), in the case (\ref{eq:SPE}), 
\[ 
 |\beta_l^{(n)}(\lambda)|=2^{-l} |\Gamma ((n+d/2-l+i\lambda)/2)\Gamma^{-1} ((n+d/2+l+i\lambda)/2)| .
\]
It follows from the identity
\begin{equation}
\Gamma(z+1)=z \Gamma(z)
\label{eq:GAM}\end{equation}
that
\[ 
 |\beta_l^{(n+2)}(\lambda)|=  |n+d/2-l+i\lambda| |n+d/2+l+i\lambda|^{-1}\: |\beta_l^{(n)}(\lambda)|
\leq |\beta_l^{(n)}(\lambda)|.
\]
Therefore the numbers
\[ 
   \sigma_l^{(n)}=\min_{\lambda\in{\Bbb R}}|\beta_l^{(n)}(\lambda)|^{-1}
\]
are related by the inequality
$ \sigma_l^{(n)}\leq    \sigma_l^{(n+2)}$.
In particular, we obtain the following result.

\begin{lemma}\label{REL}
For any $m=0,1,2,\ldots$
\[
\sigma_l^{(2m)}\geq \sigma_l^{(0)}, \quad \sigma_l^{(2m+1)}\geq \sigma_l^{(1)}.
\]
\end{lemma}

 Let us check that   the minimum in definition (\ref{eq:SPEA}) of $\sigma_l^{(0)}$ and
$\sigma_l^{(1)}$ is attained at the point
$\lambda=0$. To that end we need the following assertion from the theory of the $\Gamma$-function.

\begin{lemma}\label{gamma}  
Let $b>0$, $a\leq b$ and $\lambda\in{\Bbb R}$. Then inequality
\begin{equation} 
 |\Gamma (a+i\lambda)\Gamma^{-1} (b+i\lambda)|\leq
|\Gamma (a )\Gamma^{-1} (b)|
\label{eq:Ga1}\end{equation}
holds in the following three cases: $1^0$ $a>0$, $2^0$  $a=-n+\varepsilon$ where $n=1,  2,\ldots$,
$\varepsilon\in(0,1)$ and $\varepsilon\leq b$, $3^0$ $a\in (-1,0)$ and $|a|\leq b$.
\end{lemma}
{\it Proof.} --
Let us start with the first case. Clearly, (\ref{eq:Ga1}) is equivalent to the inequality
\[
 \Gamma (a+i\lambda) \Gamma (a-i\lambda) \Gamma (a)^{-2}\leq 
 \Gamma (b+i\lambda) \Gamma (b-i\lambda) \Gamma (b)^{-2},\quad 0<a\leq b.
\]
Thus, it suffices to check that for any  $\lambda\in{\Bbb R}$ the derivative of the function
\[
\varphi(a,\lambda)= \Gamma (a+i\lambda) \Gamma (a-i\lambda) \Gamma (a)^{-2}
\]
with respect to $a$ is nonnegative. Calculating this derivative and denoting 
$\psi(z)= \Gamma(z)^{-1} \Gamma^\prime (z)$, we find that
\[
\varphi(a,\lambda)^{-1} \partial \varphi(a,\lambda)/\partial a= \psi(a+i\lambda)+
\psi(a-i\lambda)-2\psi(a).
\]
 It follows from the Dirichlet representation (formula (20), section 1.7 of  \cite{BE}, vol. 1) 
\[
\psi(z)=\int_0^\infty (e^{-x}-(1+x)^{-z})x^{-1} dx, \quad\Re z >0,
\]
that  
\[
\psi(a+i\lambda)+\psi(a-i\lambda)-2\psi(a)=2\int _0^\infty (1-\cos(\lambda\ln(1+x)))
(1+x)^{-a} x^{-1} dx \geq 0, \quad a >0.
\]

To consider the case $2^0$, we remark that, by (\ref{eq:GAM})
\begin{equation} 
 |\Gamma (a+i\lambda) \Gamma^{-1} (b+i\lambda)| = |(-n+\varepsilon+i\lambda)^{-1}\cdots 
(-1+\varepsilon+i\lambda)^{-1}\Gamma (\varepsilon+i\lambda) \Gamma^{-1} (b+i\lambda)|.
\label{eq:Ga3}\end{equation}
Using (\ref{eq:Ga1}) for the numbers $\varepsilon$ (in place of $a$), $b$ and obvious estimates
$|(-k+\varepsilon+i\lambda)^{-1}|\leq
|(-k+\varepsilon )^{-1}|,\; k=1,\cdots, n,$
we find that
 the right-hand side of (\ref{eq:Ga3}) is bounded by 
\[
|(-n+\varepsilon )^{-1}\cdots  (-1+\varepsilon )^{-1}\Gamma (\varepsilon ) \Gamma^{-1} (b)|,
\]
which, again by    (\ref{eq:GAM}), equals $|\Gamma (a ) \Gamma^{-1} (b)|$. 

To prove the part $3^0$, we use again (\ref{eq:GAM}), apply inequality (\ref{eq:Ga1}) to the
numbers $a+1$ and $b+1$  and remark that 
\[
|b+i\lambda| |a+i\lambda|^{-1}\leq b |a|^{-1}.
\]
 This yields
\[
 |\Gamma (a+i\lambda) \Gamma ^{-1}(b+i\lambda)|\leq |\Gamma (a+1+i\lambda)\Gamma^{-1}
(b+1+i\lambda)|  |b+i\lambda| |a+i\lambda|^{-1} \leq |\Gamma (a+1) \Gamma^{-1}
(b+1)| b |a|^{-1}.
\]
The right-hand side here equals $|\Gamma (a ) \Gamma^{-1} (b)|.\quad\Box$

Now we can simplify expressions for $\sigma_l^{(0)}$ and $\sigma_l^{(1)}$.

\begin{lemma}\label{sigma}   
Put
\begin{equation}
\sigma_l^{(c)}=2^l |\Gamma ((d/2+l)/2) \Gamma^{-1} ((d/2-l)/2)|,\quad
\sigma_l^{(s)}=2^l |\Gamma ((d/2+l+1)/2) \Gamma^{-1} ((d/2-l+1)/2)|.
\label{eq:si}\end{equation}
Then $\sigma_l^{(0)}=\sigma_l^{(c)}$ and $\sigma_l^{(1)}=\sigma_l^{(s)}$.
\end{lemma}
 {\it Proof.} --
Consider first  $\sigma_l^{(1)}$.  
According to (\ref{eq:Ga1}) it suffices to check that numbers $a=(d/2-l+1)/2$ and $b=(d/2+l+1)/2$
satisfy one of the three conditions of Lemma~\ref{gamma}. If $d\geq 2$, then $b>1$ and hence
condition $2^0$ holds for all $l>0$. If $d=1$, we distinguish the cases $l>3/2$ and $l<3/2$. In the
first of them
$b>1$ so that condition $2^0$ is fulfilled and in the second $a>0$ so that condition $1^0$ is
fulfilled.
 
Similarly, to consider    $\sigma_l^{(0)}$, we need  to check
that numbers $a=(d/2-l)/2$, $b=(d/2+l)/2$ also satisfy one of the three conditions of
Lemma~\ref{gamma}. If $d\geq 4$, then $b>1$ and hence condition $2^0$ holds  for all $l>0$. If
$d=3$, we distinguish the cases $l>3/2$ and $l<3/2$. In the first of them
$b>1$ so that condition $2^0$ is fulfilled and in the second $a>0$ so that condition $1^0$ is
fulfilled.  If $d=2$, we distinguish the cases $l>1$ and $l<1$. In the first of them
$b>1$ and condition $2^0$ is fulfilled. In the second $a>0$ and condition $1^0$ is
fulfilled. Let, finally, $d=1$. If $l<1/2$, then $a=1/4-l/2>0$  so that condition $1^0$ holds. If
$l>3/2$, then $b=1/4+l/2>1$  so that condition $2^0$ holds. Finally, in the case $l\in (1/2,3/2)$
we have that $a\in (-1/2,0)$ and $|a|=l/2-1/4$. Therefore $|a|<b$ and we can refer 
to condition $3^0.\quad\Box$

{\bf 4}.
Let us return to the operators ${\bf H}_\gamma^{(c)}={\bf H}_0+\gamma {\bf V}^{(c)}$ and ${\bf
H}_\gamma^{(s)}={\bf H}_0+\gamma {\bf V}^{(s)}$ in the space
$L_2({\Bbb R}^d)$. We consider first the one-dimensional case $d=1$
when (\ref{eq:DEC}) reduces to the decomposition of the space $L_2({\Bbb R})$
into the subspaces of the even and odd functions. These subspaces are invariant with respect to
${\bf H}_\gamma^{(c)}={\bf H}_0+\gamma {\bf V}^{(c)}$ and ${\bf H}_\gamma^{(s)}={\bf H}_0+\gamma
{\bf V}^{(s)}$, ${\bf V}^{(c)}f=0$ for odd $f$ and  ${\bf V}^{(s)}f=0$ for even $f$. Therefore, the
negative spectrum of the operator
${\bf H}_\gamma^{(c)}$ (respectively, ${\bf H}_\gamma^{(s)}$) in the space
$L_2({\Bbb R})$ coincides with that of the operator $H_\gamma $ for $v(t)=(2/\pi)^{1/2}\cos t$
(respectively,
$v(t)=(2/\pi)^{1/2}\sin t$)  in the space $L_2({\Bbb R}_+)$. Thus, we can directly apply
Proposition~\ref{BES}, where according to
(\ref{eq:PC}) $  p=-1/2, q=1/2$ for the operator ${\bf H}_\gamma^{(+)}$ and  $ p=1/2, q=1/2$ for
the operator ${\bf H}_\gamma^{(-)}$. Moreover, in the case $d=1$ expressions (\ref{eq:si}) can be a
little bit simplified. This gives us the following result.

\begin{theorem}\label{d1}
Let  $d =1$. Put
\[
\sigma_l^{(c)}=(\pi/2)^{1/2}|\cos (\pi(1/2-l)/2) \Gamma (1/2-l)|^{-1},
\quad
\sigma_l^{(s)}=(\pi/2)^{1/2}|\sin (\pi(1/2-l)/2) \Gamma (1/2-l)|^{-1}.
\]

 Suppose that $l\neq 1/2+ 2k$ for any $k=0,1,2,\ldots$.
Then the negative spectrum of the operator ${\bf H}_\gamma^{(c)}$  is empty if $l\in (0, 1/2)$  and 
$|\gamma|\leq \sigma_l^{(c)}$. If  $l\in (2k+1/2, 2k+5/2)$, then it consists of
$[(k+1)/2]$ eigenvalues for  $\gamma\in (0,\sigma_l^{(c)}]$ and it consists of
$[k/2]+1$ eigenvalues for $\gamma\in [-\sigma_l^{(c)},0)$. If $|\gamma|\geq \sigma_l^{(c)}$, then,
for any $l$, the negative spectrum of the operator ${\bf H}_\gamma^{(c)}$  is infinite.

Suppose that $l\neq  3/2+2k$ for any $k=0,1,2,\ldots$. Then the negative spectrum of the operator
${\bf H}_\gamma^{(s)}$  is empty if $l\in (0, 3/2)$  and 
$|\gamma|\leq \sigma_l^{(s)}$. If  $l\in (2k+3/2, 2k+7/2)$, then it consists of
$[(k+1)/2]$ eigenvalues for  $\gamma\in (0,\sigma_l^{(s)}]$ and it consists of
$[k/2]+1$ eigenvalues for $\gamma\in [-\sigma_l^{(s)},0)$. If $|\gamma|\geq \sigma_l^{(s)}$, then,
for any $l$, the negative spectrum of the operator ${\bf H}_\gamma^{(s)}$  is infinite.
\end{theorem}

{\bf 5}.
Let now $d\geq 2$.
It follows from Proposition~\ref{Hn} and Lemmas~\ref{REL}, \ref{sigma} that in the case 
$|\gamma|\leq \sigma_l^{(c)}$ (respectively, $|\gamma|\leq \sigma_l^{(s)}$)
the negative spectra of all operators
$H_\gamma^{(2m)}$ (respectively,
$H_\gamma^{(2m+1)}$) are finite.   Moreover, they are empty for $m$ large enough. Therefore, by
virtue of (\ref{eq:DEC}), in this case the negative spectrum of the operator
${\bf H}_\gamma^{(c)}$ (respectively, ${\bf H}_\gamma^{(s)}$) is finite. On the other hand, if
$|\gamma| > \sigma_l^{(c)}$ (respectively, 
$|\gamma|> \sigma_l^{(s)}$), then the negative spectrum of the operator
$H_\gamma^{(0)}$  (respectively, $H_\gamma^{(1)}$) is infinite.
This gives us  necessary and sufficient conditions of the finiteness of
the negative spectum of these operators.

\begin{theorem}\label{FIN}
Let the numbers $\sigma_l^{(c)}$ and  $\sigma_l^{(s)}$ be defined by $(\ref{eq:si})$. Suppose that
 $l\neq d/2+ 2k$ for any $k=0,1,2,\ldots$. Then the negative spectrum of the operator
${\bf H}_\gamma^{(c)}$ is finite if and only if $|\gamma|\leq \sigma_l^{(c)}$.
Suppose that
 $l\neq d/2+ 2k+1$ for any $k=0,1,2,\ldots$. Then the negative spectrum of the operator
${\bf H}_\gamma^{(s)}$ is finite if and only if $|\gamma|\leq \sigma_l^{(s)}$.
\end{theorem}

Note that Theorems~\ref{INF} and \ref{FIN} can be unified  since
  $\sigma_l^{(c)}=0$ (respectively, $\sigma_l^{(s)}=0$) if $l = d/2+
2k$ (respectively, $l = d/2+ 2k+1$) for some $k=0,1,2,\ldots$.

It remains to calculate the total numbers ${\bf N}^{(\pm)}_{l,c}$ and ${\bf N}^{(\pm)}_{l,s}$  of
negative eigenvalues of the operators
${\bf H}_\gamma^{(c)}$ and ${\bf H}_\gamma^{(s)}$ in the cases $\pm \gamma \in(0, \sigma_l^{(c)}]$
and
$\pm \gamma \in(0, \sigma_l^{(s)}]$, respectively. We proceed from decomposition (\ref{eq:DEC})  and
rely on Proposition~\ref{Hn}. Recall also that numbers $\nu_l$ were defined by equality
(\ref{eq:ex3}).
 Consider, for example,  ${\bf H}_\gamma^{(c)}$. 
 Suppose first that $\gamma\in [-\sigma^{(c)}_l,0)$. If $l<d/2$, then
 $H_\gamma^{(2m)}\geq 0$  for all $m$ so that ${\bf N}^{(-)}_{l,c}=0$. If $l\in (d/2,d/2+2)$,
then the operator
$H_\gamma^{(0)}$ has one negative eigenvalue and 
$H_\gamma^{(2m)}\geq 0$  for $m\geq 1$. Since $\nu_0=1$, in this case ${\bf N}^{(-)}_{l,c}=1$. If
$l\in (d/2+2,d/2+4)$, then the operator
$H_\gamma^{(0)}$ has two negative eigenvalues and 
$H_\gamma^{(2m)}\geq 0$  for $m\geq 1$ so that ${\bf N}^{(-)}_{l,c}=2$.  If $l\in (d/2+4,d/2+6)$,
then the operator
$H_\gamma^{(0)}$ has three negative eigenvalues, the operators
$H_\gamma^{(2)}$ and $H_\gamma^{(4)}$ have one negative eigenvalue each and 
$H_\gamma^{(2m)}\geq 0$  for $m\geq 3$. It follows that in this case ${\bf N}^{(-)}_{l,c}=3+
\nu_2+\nu_4$. Repeating this procedure, we arrive at the general formula for the case $l\in
(d/2+2k,d/2+2k+2)$:
\begin{equation}
{\bf N}^{(-)}_{l,c} =k+1+\sum_{p=0}^{[(k-1)/2]} (k-2p-1) (\nu_{4p+2}+ \nu_{4p+4}),\quad k\geq 1. 
\label{eq:NN1}\end{equation}

 The case $\gamma\in (0,\sigma^{(c)}_l]$ can be studied quite similarly. Now 
$H_\gamma^{(2m)}\geq 0$   for all $m$ if $l<d/2+2$ and hence ${\bf N}^{(+)}_{l,c}=0$ for such $l$.
If $l\in (d/2+2,d/2+4)$, then  the operators
$H_\gamma^{(0)}$ and $H_\gamma^{(2)}$ have one negative eigenvalue each and  
$H_\gamma^{(2m)}\geq 0$  for $m\geq 2$. It follows that in this case ${\bf N}^{(+)}_{l,c}=
\nu_0+\nu_2$. If $l\in (d/2+4,d/2+6)$, then  the operators
$H_\gamma^{(0)}$ and $H_\gamma^{(2)}$ have two negative eigenvalues each and again
$H_\gamma^{(2m)}\geq 0$  for $m\geq 2$ so that ${\bf N}^{(+)}_{l,c}= 2(\nu_0+\nu_2)$.
If $l\in (d/2+6,d/2+8)$, then  the operators
$H_\gamma^{(0)}$ and $H_\gamma^{(2)}$ have three negative eigenvalues each, both operators
$H_\gamma^{(4)}$ and $H_\gamma^{(6)}$ have exactly one negative eigenvalue  and  
$H_\gamma^{(2m)}\geq 0$  for $m\geq 4$. In this case ${\bf N}^{(+)}_{l,c}= 3(\nu_0+\nu_2) +
(\nu_4+\nu_6)$. The general formula for the case $l\in (d/2+2k,d/2+2k+2)$ reads as
\begin{equation}
 {\bf N}^{(+)}_{l,c} = \sum_{p=0}^{[k/2]} (k-2p) (\nu_{4p }+ \nu_{4p+2}). 
\label{eq:NN2}\end{equation}
Let us formulate these results.

\begin{theorem}\label{+}
  The number ${\bf N}^{(-)}_{l,c}=0$ for $l<d/2$, ${\bf N}^{(-)}_{l,c}=1$ for $l\in
(d/2,d/2+2)$ and   ${\bf N}^{(-)}_{l,c}$   is determined by formula $(\ref{eq:NN1})$
for $l\in (d/2+2k,d/2+2k+2)$.
 The number ${\bf N}^{(+)}_{l,c}=0$ for $l<d/2+2$
and   ${\bf N}^{(+)}_{l,c}$    is determined by formula $(\ref{eq:NN2})$
for $l\in (d/2+2k,d/2+2k+2)$.
\end{theorem}

The total numbers ${\bf N}^{(\pm)}_{l,s}$  of negative eigenvalues
of the operator ${\bf H}_\gamma^{(s)}$ can be found quite similarly.

\begin{theorem}\label{-}
 The number ${\bf N}^{(+)}_{l,s}=0$  for $l<d/2+1$,
${\bf N}^{(+)}_{l,s}=\nu_1$ for $l \in (d/2+1,d/2+3)$ and 
\[ {\bf N}^{(+)}_{l,s} =(k+1)\nu_1+ \sum_{p=1}^{[(k+1)/2]} (k-2p +1) (\nu_{4p-1}+ \nu_{4p+1}) 
\]
 for $l\in (d/2+1+2k,d/2+3+2k),\; k\geq 1$.
 The number ${\bf N}^{(-)}_{l,s}=0$ for $l<d/2+3$ and  
\[
{\bf N}^{(-)}_{l,s} = \sum_{p=0}^{[k/2]} (k-2p ) (\nu_{4p+1}+ \nu_{4p+3}). 
\]
 for $l\in (d/2+1+2k,d/2+3+2k)$.
\end{theorem}

We emphasize that the  numbers ${\bf N}^{(\pm)}_{l,c}$ and ${\bf N}^{(\pm)}_{l,s}$ do not depend on
coupling constants $\pm \gamma \in(0, \sigma_l^{(c)}]$ and
$\pm \gamma \in(0, \sigma_l^{(s)}]$, respectively.

\end{document}